\documentclass[twocolumn,showpacs,amsmath,amssymb,aps,10pt,reprint]{revtex4-1}
\usepackage{graphics}
\usepackage{amsmath}
\usepackage{amssymb}
\usepackage[breaklinks=true,colorlinks=true,linkcolor=blue,urlcolor=blue,citecolor=blue]{hyperref}
\usepackage{soul,color}
\usepackage{epstopdf}
\usepackage{balance}
\usepackage{multirow}

\usepackage{rotating}

\begin{document}

\title{Pinning effects on flux flow instability in epitaxial Nb thin films}

\author{
Oleksandr~V.~Dobrovolskiy$^{1}$,
Valerij~A.~Shklovskij$^{2}$,\\
Marc Hanefeld$^1$,
Markus Z\"orb$^1$,
Lukas K\"ohs$^1$, and
Michael Huth$^1$
}

\address
        {
        $^1$Physikalisches Institut, Goethe University, 60438 Frankfurt am Main, Germany\\
        $^2$Physics Department, V. N. Karazin Kharkiv National University, 61077 Kharkiv, Ukraine
        }

\date{\today}

\begin{abstract}
The flux flow properties of epitaxial niobium films with different pinning strengths are investigated by dc electrical resistance measurements and mapped to  results derived within the framework of a theoretical model. Investigated are the cases of weak random pinning in as-grown films, strong random pinning in Ga ion-irradiated films, and strong periodic pinning induced by a nanogroove array milled by focused ion beam. The generic feature of the current-voltage curves of the films consists in instability jumps to the normal state at some instability current density $j^\ast$ as the vortex lattice reaches its critical velocity $v^\ast$. While $v^\ast(B)$ monotonically decreases for as-grown films, the irradiated films exhibit a non-monotonic dependence $v^\ast(B)$ attaining a maximum in the low-field range. In the case of nanopatterned films, this broad maximum is accompanied by a much sharper maximum in both, $v^\ast(B)$ and $j^\ast(B)$, which we attribute to the commensurability effect when the spacing between the vortex rows coincides with the location of the grooves. We argue that the observed behavior of $v^\ast(B)$ can be explained by the pinning effect on the vortex flow instability and support our claims by fitting the experimental data to theoretical expressions derived within a model accounting for the field dependence of the depinning current density.
\end{abstract}


\maketitle
\section{Introduction}
The vast majority of technologically important superconductors are superconductors of type II: Magnetic field penetrates these as a flux-line array of Abrikosov vortices. The repulsive interaction between vortices makes them to arrange in most cases into a triangular lattice. In the presence of a transport current density $j$ which is smaller than the depinning current $j_d$, the vortex lattice is pinned. In the opposite case $j > j_d$, i.\,e. when the Lorentz force exerted by the transport current dominates the pinning force, the vortex ensemble moves causing dissipation associated with heat generation. As a well-known approach, an enhancement of $j_d$ leading to a reduction of dissipation can be achieved by strategically introducing vortex pinning sites \cite{Wor17boo,Dob17pcs}. However, to only enhance $j_d$ is not enough to preserve the low-resistive flux-flow state: At high vortex velocities, flux flow becomes instable \cite{Lar76etp}, leading to an abrupt transition to the normal state clearly visible in the electric field (E) --- current density (j) curve (CVC). This transition happens at some instability velocity $v^\ast$ of the vortices corresponding to the instability current density $j^\ast$ which is notably smaller than the Ginzburg-Landau depairing current density $j_{GL}$. Accordingly, finding strategies for expanding the low-dissipative current-carrying capability of superconductors into the regime of high current densities by enhancing the ``speed limits'' of the vortex dynamics is a topical problem which has been attracting significant interest in the last decades \cite{Xia99prb,Siv03prl,Sil12njp,Gri15prb}.

There are several flux-flow instability mechanisms discussed to explain voltage jumps in the CVCs. First, the instability may be due to a thermal runaway effect  due to Joule heating \cite{Xia99prb,Gon03prb}. In this case, high current densities induce a power dissipation in the film that is enough to destroy superconductivity. The temperature of the sample abruptly rises and the fingerprint of this mechanism is that the dissipated power at the instability point is independent of the magnetic field. Second, a hot-spot effect may occur consisting in the formation of localized normal domains which appear in places of maximum current due to an inhomogeneous current distribution and are sustained by Joule heating \cite{Vod07prb,Sil10prl}. Similar to the previous case, the resulting CVC is characterized by a hysteresis when ramping current up and down. Third, the vortex system can undergo crystallization \cite{Kos94prl,Kok04prb}. This transition is known to occur at close-to-critical currents, between a pinned state and a coherently moving lattice at large velocities. Fourth, if caused by phase-slip centers or lines \cite{Siv03prl,Ber09prb} the instability manifests itself in the CVC as a voltage-step structure, with segments of constant dynamic resistance exhibiting a field-independent slope. Fifth, and perhaps the most extensively discussed, is the Larkin-Ovchinnikov (LO) mechanism~\cite{Lar86inb} close to $T\simeq T_c$ and later on refined by Bezuglyi and Shklovskij (BS) \cite{Bez92pcs}: The electric field generated by the vortex motion shifts the distribution of quasiparticles inside the vortex core to higher energies, causing some of them to leave the core. The vortex core therefore shrinks with increasing flux-line velocity $v$, and the viscous damping coefficient $\eta$ decreases. Finally, the Kunchur hot-electron instability \cite{Kun02prl,Bab04prb} has to be mentioned, which is observed at $T \lesssim 0.5T_c$ and related to thermal effects diminishing the superconducting order leading to an expansion of the vortex cores rather than shrinkage.

Although the flux flow instability has been known since many years \cite{Mus80etp,Kle85ltp,Doe94prl,Per05prb,Xia96prb}, the effect of pinning on the instability parameters $v^\ast$ and $j^\ast$ has attracted interest only recently \cite{Sil12njp,Gri12apl,Gri15prb}. In particular, Grimaldi \emph{et al} \cite{Gri12apl} have shown for a series of superconductors, that a non-monotonic dependence $v^\ast(B)$ results either at moderately strong pinning or at moderate temperature, while $v^\ast(B)$ monotonically decreases at low and high temperatures and in the limits of weak and strong pinning. Here the temperatures are related to temperatures of the mixed state. There is still a significant lack of comparing experimental results to theoretical models, since the models developed so far have been widely ignoring disorder and pinning. 

To bridge this gap, the effect of pinning on the hot electron flux-flow instability has been analyzed by one of us \cite{Shk17arx} by accounting for the magnetic field dependence of the depinning current that has allowed for an explanation of the non-monotonic dependence of $v^\ast(B)$ at low fields. There, the pinning has been introduced phenomenologically by using the nonlinear conductivity generated by the a washboard pinning potential instead of the Bardeen-Stephen flux-flow conductivity in the CVC. A heat balance equation for electrons in low-$T_c$ superconducting films has been solved in Ref. \cite{Shk17arx} in the two-fluid approach. A theoretical analysis has revealed \cite{Shk17arx} that the $B$-behavior of $E^\ast$, $j^\ast$ and $\rho^\ast$ is monotonic, whereas the $B$-dependence of $v^\ast$ is quite different as $dv^\ast/dB$ may change its sign, as sometimes observed in experiments \cite{Gri12apl,Sil12njp,Gri10prb}. 

At $T\lesssim T_c$, pinning effects on the flux-flow instability parameters have recently been analyzed theoretically in Ref. \cite{Shk17snd} on the basis of the generalized LO and BS approaches. In that work \cite{Shk17snd}, it has been shown that with increasing pinning strength at a fixed magnetic field value the critical instability velocity $v^\ast$ decreases, the instability current density $j^\ast$ increases, while the dissipated power $P^\ast$ and the quasiparticles temperature $T^\ast$ remain practically constant.

In this paper, we experimentally explore the vortex dynamics in the regime of high vortex velocities in epitaxial Nb films with different pinning types and its strength at $T = 0.4T_c$, where the hot-electron mechanism of the hot-electron vortex-flow instability dominates \cite{Kun02prl}. In particular, we observe that the magnetic field dependence of the depinning current density $j_d(B)$ can be fitted to an expression of the general form $j_d \propto 1 / B^m$. However, the exponent $m \simeq 1$ is larger in Nb films with stronger pinning, i.\,e. for ion-irradiated and nanopatterned films, while $m \simeq 0.5$ for as-grown films. An account for this exponent change within the framework of the theoretical model introduced in \cite{Shk17arx} allows us to fit the crossover from the monotonic decrease of $v^\ast(B)$ in the case of the as-grown films to the non-monotonic behavior of $v^\ast(B)$ for the films with stronger pinning. In addition, we observe a sharp peak in $v^\ast(B)$ for nanostructured films, which is caused by the onset of the coherent vortex motion under fundamental matching field conditions.

\section{Experiment}
\subsection{Samples}
\begin{figure}[b!]
    \centering
    \includegraphics[width=0.8\linewidth]{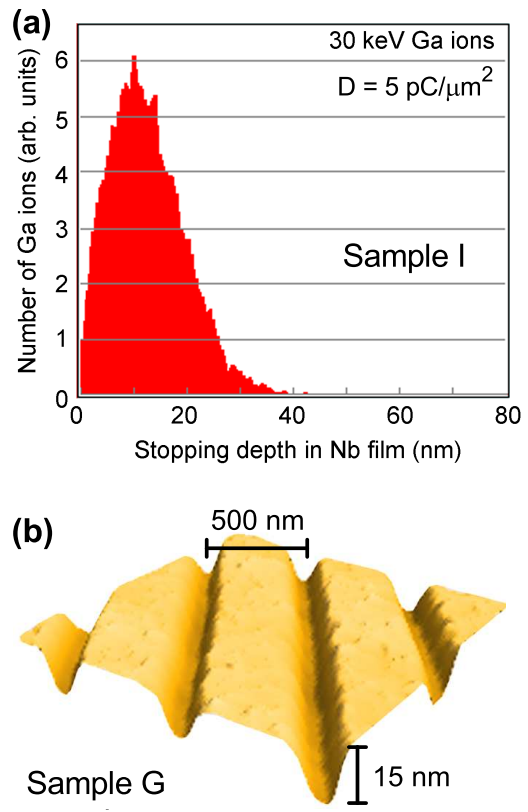}
    \caption{(a) Distribution of the implanted Ga ions with an energy of $30$\,keV over the film thickness (Sample I), simulated using the SRIM software \cite{Srim}. (b) Atomic force microscope image of the nanogroove array on the surface of the Nb film (Sample G) inducing a periodic pinning potential of the washboard type.}
    \label{fSamples}
\end{figure}

The samples are three 80\,nm-thick epitaxial (110) Nb films patterned by conventional photolithography into four-contact bridges. The films were deposited by dc magnetron sputtering with a sputtering rate of $1$\,nm/s in an ultra-high vacuum setup. The substrates were epitaxially-polished a-plane (11$\underline{2}$0) sapphire substrates kept at $850^\circ$C during the deposition process. Details of the film growth and the morphology characterization of the as-grown films are given in \cite{Dob12tsf}; we refer to film C in \cite{Dob12tsf}. One bridge was left as-deposited (sample A), the second one was irradiated by Ga ions with an accelerating voltage of $30$\,kV over the entire bridge area (sample I) with an irradiation dose of $D \simeq 5$\,pC/$\mu$m$^2$, and the third one was nanopatterned by focused ion beam (sample G). The nanopattern in sample G consisted of uniaxilly arranged 500\,nm-spaced grooves with a depth of $15$\,nm and a full width at half depth of $60$\,nm. The distribution of the Ga ions over the film thickness, simulated using the SRIM software \cite{Srim}, is shown in Fig. \ref{fSamples}(a). An atomic force microscope image the surface of the nanopatterned sample G is presented in Fig. \ref{fSamples}(b).

The as-grown films are characterized by a superconducting transition temperature $T_c = 9.1$\,K defined by using a $50$\% resistance criterion. After the exposure of the films to the ion beam, $T_c$ dropped to $8.97$\,K and $9.03$\,K for sample I and sample G, respectively. The upper critical field $H_{c2}(0)$ of all samples is about $1.1$\,T, as deduced from fitting the dependence $H_{c2}(T)$ to the phenomenological law $H_{c2}(T) = H_{c2}(0) [1-(T/T_c)^2]$.

The electrical resistance measurements were done with magnetic field oriented perpendicular to the film surface. The nanogrooves in sample G were aligned along the current flow such that the vortex motion under the action of the Lorentz force took place across the grooves. Since unavoidable self-heating can affect the experimental data, several measures have been undertaken. To minimize self-heating effects, CVCs were measured in a pulsed current-driven mode with a rectangular pulse width of 1.1\,ms and a pulse-off time of 1\,s. By sweeping the current forth and back, we verified that no hysteresis occurred in the CVCs. The instability points in each curve have been proven to remain unchanged so that this allowed us to conclude that Joule self-heating effects can be neglected in our measurements.
\begin{figure}[t!]
    \centering
    \includegraphics[width=1\linewidth]{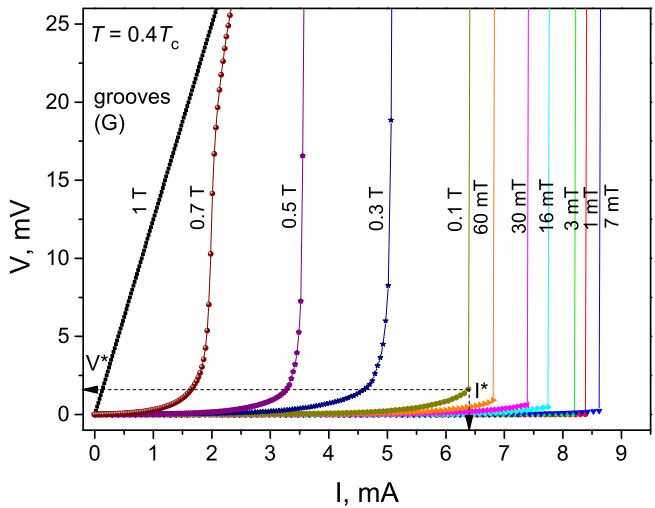}
    \caption{Current-voltage curves $V(I)$ at $T = 0.4T_c$ for sample G (with nanogrooves) for a series of selected magnetic field values, as indicated. Dashed arrows illustrate the deduction of the instability current $I^\ast$ and the instability voltage $V^\ast$ at $H = 0.1\,T$, from which the instability current density $j^\ast$ and the instability velocity $v^\ast$ are deduced.}
    \label{fCVC}
\end{figure}


The CVCs were measured at $T = 0.4T_c$ for a series of magnetic fields in the range $0$ to $1$\,T. Figure \ref{fCVC} illustrates an exemplary series of CVCs for the film with nanogrooves. For all samples the CVCs exhibit a dissipation-free regime at small current densities and a nearly linear regime relating to viscous flux flow at current densities larger than the depinning current density $j_d$. At small and moderate magnetic fields, for larger current densities the curves start to bend upward as soon as the critical voltage $V^\ast$ is reached. From the last data point before the jump at $j^\ast$, refer to Fig. \ref{fCVC}(a), the critical vortex velocity $v^\ast$ is derived by the relation
\begin{equation}
v^\ast = \frac{V^\ast}{B L},
\end{equation}
where $B$ ist the applied magnetic field and $L=100\,\mu$m ist the distance between the voltage contacts. At larger magnetic fields $B>0.3$\,T, the CVCs become more smeared and finally the instability jumps disappear and a continuous nonlinear transition to the normal state takes place. From the measured $V(I)$ data we deduce the values of the depinning current density $j_d =I_d/wd$ by using a $10$\,nV voltage criterion. Here $w = 10\,\mu$m is the sample width and $d$ is the sample thickness. The instability current density $j^\ast =I^\ast/wd$ is determined from the current $I^\ast$ relating to $V^\ast$.

\subsection{Results}
\begin{figure}[b!]
    \centering
    \includegraphics[width=1.01\linewidth]{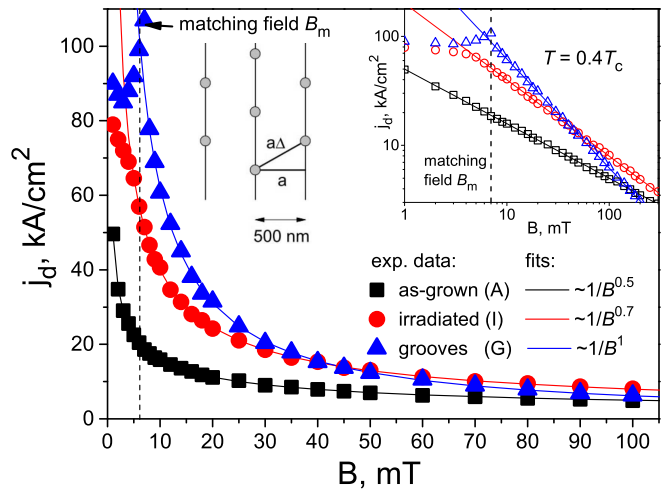}
    \caption{Symbols: Magnetic field dependence of the depinning current density $j_d(B)$ for all samples. Solid lines are fits of the general form $j_d \propto1/B^m$ with the exponent $m$ as indicated in the plot. Right inset: A log-log plot of the main panel. Left inset: The arrangement of vortices with respect to the grooves in sample G at the matching field $B_m=7.2$\,mT.}
    \label{fjvB}
\end{figure}

The magnetic field dependences of the depinning current density for all films are plotted in Fig. \ref{fjvB}. For samples A and I, the curves $j_d(B)$ are smooth decreasing functions of $B$. Sample G is characterized by a a factor of two larger depinning current densities as compared to the as-grown sample A.
This can be understood as a consequence of the pinning enhancement owing to the grooves. In addition, the $j_d(B)$ curve of sample G shows a maximum at $B= 7$\,mT. From our previous work we know that the assumed triangular vortex lattice matches with the 500\,nm-periodic nanolandscape at the fundamental matching field $B = B_m = 7.2\,$mT \cite{Dob15apl,Dob15met}, see the inset to Fig. \ref{fjvB}. This allows us to conclude that the observed peak in $j_d(B)$ is caused by the efficient enhancement of pinning at the fundamental matching condition. At this field value all vortex rows are pinned at the groove bottoms and there neither vacant grooves nor interstitial (non-pinned) vortices.

\begin{figure}[t!]
    \centering
    \includegraphics[width=1\linewidth]{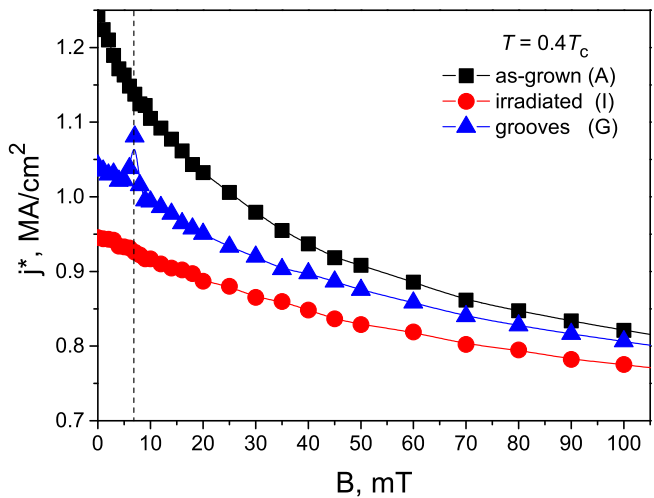}
    \caption{Symbols: Magnetic field dependence of the instability current density $j^\ast(B)$ for all samples. Solid lines are guides to the eye.}
    \label{fjinstvB}
\end{figure}

The magnetic field dependences of the instability current density $j^\ast(B)$ for all films are presented in Fig. \ref{fjinstvB}. All curves show $j^\ast(B)$ to be a decreasing function of $B$. For film G, one can also recognize a small peak in the dependence $j^\ast(B)$ near $B_m = 7$\,mT as reminiscent of the more pronounced peak in the dependence $j_d(B)$ in Fig. \ref{fjvB}. At larger magnetic fields the current range of nearly linear flux flow shrinks as $j^\ast(B)$ tends to coincide with $j_d(B)$ and, at magnetic fields $B\gtrsim 0.8$\,T (not shown), $j^\ast(B)$ can not be defined anymore as the samples transit to the normal state without instability jumps.

\begin{figure}[t!]
    \centering
    \includegraphics[width=1\linewidth]{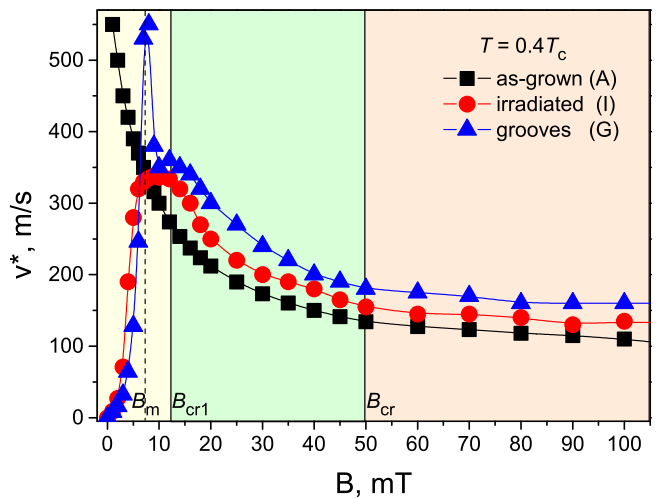}
    \caption{Magnetic field dependence of the instability velocity $v^\ast(B)$ for all samples. Solid lines are guides to the eye. The lower crossover field $B_{cr1} \approx12$\,mT is indicated for sample G. For sample I $B_{cr1} \approx 10$\,mT.}
    \label{fvvB}
\end{figure}

The magnetic field dependence of the instability velocity deduced from the CVCs for all films is presented in Fig. \ref{fvvB}. The behavior of the curves $v^\ast(B)$ for the three samples differs substantially. In particular, for sample A, $v^\ast(B)$ decreases monotonically, while the behavior of $v^\ast(B)$ for sample I and sample G is non-monotonous. For sample I the instability velocity grows with the $B$ field value at low fields and attains a maximum at the first crossover field $B_{cr1} \backsimeq 10$\,mT. With further increasing field $v^\ast(B)$ decreases as $1/\sqrt{B}$ and at fields larger than $B_{cr} \backsimeq 50$\,mT it saturates at about $150$\,m/s. In addition to the relatively broad peak at $B = B_{cr}$, the dependence $v^\ast(B)$ of sample G shows a very sharp peak at  the matching field $B\thickapprox B_{m} = 7.2$\,mT. At the matching field, the vortex instability velocity is a factor of 1.7 higher than the maximal vortex velocity achieved at $B_{cr1} \backsimeq 10$\,mT.

\section{Discussion}
\subsection{Critical velocity}
Since both, the monotonic and the non-monotonic dependences of $v^\ast(B)$ have previously been observed in different temperature ranges and for different degrees of disorder \cite{Mus80etp,Kle85ltp,Doe94prl,Per05prb,Xia96prb,Sil12njp,Gri12apl,Gri15prb}, we begin with a general discussion of the pinning mechanisms at work in our samples. In general, pinning is niobium films is known to be strong \cite{Per05prb,Gri10prb}. However, the high structural quality (epitaxial growth) of our as-deposited Nb films ensures that their intrinsic pinning is relatively weak. Indeed, the depinning current densities in sample A are by one to two orders of magnitude smaller than typical values of $j_d$ reported for Nb films \cite{Per05prb,Gri10prb}. The irradiation of the entire film area by Ga ions in sample I leads to stopping and incorporation of the ions within, largely, the 40\,nm-thick topmost region of the film \cite{Srim}, refer to Fig. \ref{fSamples}(a). Laterally, the  distribution of the Ga ions in the film is random and uniform. Due to the implanted Ga ions Sample I is characterized by a higher degree of \emph{random} disorder than sample A. This is reflected in the values of the depinning current density $j_d$, which are a factor of two higher in sample I as compared to sample A. This is in contradistinction to sample G, which has intact areas of the as-grown film between the grooves and only the grooves behave as very strong linearly-extended pins \cite{Dob12njp,Dob16sst}. We note that while the value of $j_d$ strongly depends on the density and type of pinning sites in the sample, the instability current density $j^\ast$ only showed  about $20\%$ variation over all samples in our experiment.

The degree of disorder and the different origin of pinning in different samples hence allow us to explain the observed behavior of $v^\ast(B)$ as follows. At larger fields, when the vortex lattice parameter $a = (2\Phi/\sqrt{3}B)^{1/2}$ is smaller than the quasiparticle diffusion length, the velocity $v^\ast$ is independent of $B$, which is consistent with the LO prediction for higher temperatures ($T \simeq T_c$). With decreasing $B$, the vortex lattice parameter is increasing and becomes comparable with the quasiparticle diffusion length such that the instability velocity related to $a$ via the quasiparticle scattering time $\tau_E$ as $v^\ast \tau_E \sim a$ increases as $1/\sqrt{B}$ with decreasing field at fields $B < B_{cr}$. Hence, from this crossover field one can estimate the quasiparticle scattering time as $\tau_E \simeq10$\,ns in our Nb films. With further reduction of the vortex density at lower magnetic fields, the pinning efficiency grows since the now small number of vortices can be pinned to even few pinning sites. In this way, the random pinning disturbs the moving vortex lattice which, at the same time, exhibits softening at smaller fields. In consequence of this, there appear vortices moving more slowly and faster than the mean velocity of the vortex ensemble. The flux-flow instability takes place as soon as the instability velocity is reached by ``faster'' vortices. Since the distribution of vortex velocities becomes broadened and the difference between the mean vortex velocity and the velocity of ``faster vortices'' is growing with decreasing magnetic field, this results in the observed decreasing $v^\ast(B)$ behavior with decreasing field below $B_{cr1}$.

According to the pinning picture above, the absence of a low-field crossover in $v^\ast(B)$ for sample A can be understood as a consequence of the fact that the random pinning in this sample remains too weak even for the softened vortex lattice. The increased strength of the random pinning in sample I causes the vortex lattice to become soft at notably higher fields, such that the mentioned crossover at $B_{c1}$ is observed. Finally, in the nanostructured sample G an additional sharp peak to the left of the crossover field $B_{cr1}$ can be attributed to the enhanced coherence of the moving vortex lattice at the matching field. This coherence becomes apparent as Shapiro steps \cite{Dob15snm,Dob15mst} in independent CVC measurements in the presence of combined dc and high-frequency ac stimuli. When the vortex motion is coherent, the deviation of the velocity of individual vortices from the average vortex velocity corresponding to the velocity of the whole vortex ensemble is smallest. This means that the velocity distribution is most sharp and in the CVC the vortex motion remains stable up to higher velocities as compared to the non-matching fields. Out of matching, the coherence in the vortex dynamics is lost, as there are vortices whose instantaneous velocity is noticeably larger or smaller than the average velocity of the vortex ensemble. Hence, the distribution of vortex velocities is broader and at already smaller velocity values the ``high-velocity'' part of this distribution attains the critical velocity value that leads to an avalanche-like transition to the normal state.

\subsection{Pinning effect on flux-flow instability}
For a quantitative interpretation of the experimental results we use a phenomenological model\cite{Shk17arx} for the hot-electron instability which accounts for the pinning effect on it. In this model, the effect of pinning is introduced via the parameter $\mu$ defined as
\begin{equation}
\label{e2mu}
    2\mu \equiv (j_d/j_0^\ast)^2.
\end{equation}
where $j_d$ is the depinning current density and $j_0^\ast$ is the instability current density in the film without pinning. The limiting case of no pinning with $\mu = 0$ is realized when $j_d = 0$, while the upper limit for the depinning current $j_d = j^\ast_0$ results in $\mu = 0.5$.

The theory further relies upon a magnetic field dependence of $\mu$ assuming the following scaling for the dependence of the depinning current density on the magnetic field
\begin{equation}
\label{eScaling}
        j_d(B) = j_B(B_c/B)^m,
\end{equation}
where $j_B$ and $B_c$ are constants deduced from experimental values of $j_d(B)$, while the exponent $m>0$ is the main parameter which determines the $B$-dependence of $\mu$. It is this parameter which must be determined from experiment. Obviously, with $m=0$ Eq.~\eqref{eScaling} yields the $B$-independent case $j_d = const$, while for $m=1/2$ one has $\mu(B)=const$ since
\begin{equation}
\label{eMuvB}
        \mu(B) = j_d^2(B)B/2\gamma^2 = KB^{1-2m},
\end{equation}
where $K = j_B^2B_c^2/2\gamma^2$ and $\gamma$ is a constant. Further, the magnetic field behavior of the critical velocity is predicted to be primarily determined by the derivative of the parameter $\mu$ with respect to magnetic field \cite{Shk17arx}
\begin{equation}
\label{edMdB}
        d\mu(B)/dB = (1-2m)\mu(B)/B.
\end{equation}
One can see that $d\mu/dB$ \emph{changes its sign} at $m=1/2$. Namely, $\mu(B)$ decreases with increasing $B$ for $m>1/2$, whereas $\mu(B)$ increases for $0<m<1/2$.

In Ref. \cite{Shk17arx}, a heat balance equation for the relaxation of ``hot electrons'' \cite{Kun02prl} has been solved in conjunction with the extremum condition for the CVC derivative at the instability point. The pinning was introduced into the CVC by using the nonlinear conductivity calculated at $T = 0$ for a saw-tooth \cite{Shk06prb} and a cosine \cite{Shk08prb} washboard pinning potential instead of the Bardeen-Stephen flux-flow conductivity \cite{Kun02prl}. We note that neither the CVCs themselves nor the magnetic-field dependences of the instability critical parameters for different pinning strengths allow a linear transformation to fit one another, as their dependences are strongly nonlinear (and non-trivial) rather than of some ``universal behavior''.
In result of the carried out analysis \cite{Shk17arx}, it has been revealed that a particular form of the pinning potential does not affect the considered physics. Thus, for a superconducting film with a cosine washboard pinning potential the following expression has been derived for the instability velocity
\begin{equation}
\label{eVastB}
        v^\ast(B) = \frac{v^\ast_{max}}{\sqrt{B}}\frac{1}{\sqrt{\sqrt{1+\mu^2} + \mu}},
\end{equation}
where $v^\ast_{max} $ is a fitting parameter.

The structure of Eq. (\ref{eVastB}) can be commented as follows. The expression under the square root in the denominator of Eq. (\ref{eVastB}) follows from the solution of the heat balance equation at the instability point. The pinning effect on the instability critical velocity is mediated by the dependence of the parameter $\mu$ on $B$, thus incorporating the dependence of the depinning current density on the magnetic field value. In the limiting case of no pinning $\mu = 0$ one naturally returns to the result of Kunchur \cite{Kun02prl}
\begin{equation}
\label{eVastBK}
        v^\ast(B) \propto \frac{1}{\sqrt{B}}.
\end{equation}

In the general case of arbitrary pinning strengths, it has been shown \cite{Shk17arx} that while $v^\ast(B)$ decreases with increasing magnetic field when $m<1/2$, for $m>1/2$ the derivative $dv^\ast/dB$ should be analyzed
\begin{equation}
\label{edVAstdB}
        \frac{dv^\ast}{dB} = -A\frac{\left[1 +\displaystyle\frac{(1-2m)\mu}{\sqrt{1+\mu^2}}\right]}{B\sqrt{B}\sqrt{\sqrt{1+\mu^2} + \mu}},
\end{equation}
where $A$ is a constant. It has been revealed \cite{Shk17arx} that for $m>1/2$ the expression in the square brackets in Eq. \eqref{edVAstdB} may be negative at $m>(1 + \sqrt{1+1/\mu^2})/2$. In particular, for $\mu\gtrsim 2$ one has $m\simeq 1 + 1/(2\mu)^2$. Obviously, a consequence of the negative bracket is $dv^\ast/dB > 0$ for $B\rightarrow 0$. Accordingly, when this is the case, $v^\ast(B)$ grows with increasing field in the low-field range, attains a maximum at some field value and then decreases at higher fields.

Proceeding to a quantitative analysis of the experimental data, we first fit the curves $j_d(B)$ in Fig. \ref{fjvB} to Eq. \eqref{eScaling} with $j_B$ and $m$ as fitting parameters and $B_{c} = B_{c2}(0.4T_c) = 1\,$T as the (upper) critical field at $T = 0.4T_c$. The best fits are obtained with $m=0.5$ for film A, $m= 0.7$ for film I, and $m= 1$ for film G, as depicted by solid lines in Fig. \ref{fjvB}.

\begin{figure}[t!]
    \centering
    \includegraphics[width=0.95\linewidth]{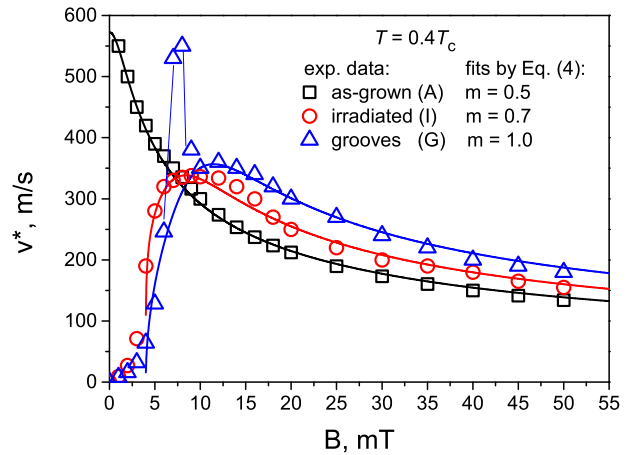}
    \caption{Experimental $v^\ast(B)$ dependences (symbols) and theoretical fits to Eq. \eqref{eVastB} (solid lines) for scaling exponents $m$ as indicated. The thin solid line connecting the symbols at the matching peak at $B = 7.2$\,mT is guide to the eye.}
    \label{fvfits}
\end{figure}

Next, we use the deduced exponents to fit the dependences of the instability velocity on the magnetic field $v^\ast(B)$ by using Eq. \eqref{eVastB}. The resulting theoretical curves (solid lines) are shown in Fig. \ref{fvfits} along with the experimental data (symbols). From the figure follows that the theoretical curve fits nicely the behavior of $v^\ast(B)$ for sample A, for which the law $v^\ast(B) \propto 1/B^{0.5}$ is observed in the full range from 0 to 50\,mT. For sample I and sample G the theoretical curves are good fits with exception of the very-low-field range from 0 to 4\,mT. We attribute this to peculiarities of the magnetic flux penetration into the samples \cite{Gri10prb}, which are not incorporated in our model \cite{Shk17arx}. Finally, Eq. \eqref{eVastB} describes very well the overall behavior of $v^\ast(B)$ for sample G with the exception of the field range in the vicinity of the matching field $B = 7.2$\,mT.

In all, the use of the scaling exponents $m$ deduced from the experimental dependences of the depinning current $j_d(B)$ on the magnetic field has allowed us to successfully fit the dependences $v^\ast(B)$ for the three samples with different pinning types.

\section{Conclusion}

To summarize, the flux-flow instability has been investigated in as-grown, ion-irradiated and nanopatterned epitaxial Nb films. The instability current densities $j^\ast$ have been revealed to only weakly depend on the pinning strength of the films, which is in contrast to the depinning current being very sensitive to the distribution and type of pinning sites in the samples. Both, irradiated and nanopatterned Nb films demonstrate a non-monotonic dependence of the instability velocity $v^\ast$ on the magnetic field $B$, attaining a maximum at a crossover field $B_{cr1}$. In addition to the broad maxima of $v^\ast(B)$ at $B_{cr1}$, $v^\ast(B)$ exhibits a pronounced sharp maximum at the fundamental matching field when the vortex lattice is commensurate with the periodic pinning structures. A change of the exponent $m$ in the dependence of the depinning current on the magnetic field $j_d(B) \propto 1/B^m$ has been observed, with $m = 0.5$ for the as-grown film, $m = 0.7$ for the ion-irradiated film, and $m=1$ for the film with periodic nanogrooves. A theoretical account for this exponent change has allowed us to derive an analytical expression for $v^\ast(B)$ and to quantitatively describe the different behaviors of $v^\ast(B)$ observed in our experiments.

\section*{Acknowledgements}
The authors thank R. Sachser for helping with nanopatterning and automating the data acquisition. This work received funding through DFG project DO1511/3-1. Further, funding from the European Unions Horizon 2020 research and innovation program under Marie Sklodowska-Curie Grant Agreement No. 644348 (MagIC) is acknowledged.


%

\end{document}